\documentclass[11pt,preprint]{aastex}
\usepackage{amssymb}
\usepackage{epsfig}

\newcommand{\kms}{\,{\rm km\,s^{-1}}}

\newcommand{\yr}{\,{\rm yr}}

\newcommand{\pc}{\,{\rm pc}}
\newcommand{\kpc}{\,{\rm kpc}}	
\newcommand{\mpc}{\,{\rm Mpc}}

\newcommand{\msun}{\,M_\odot}

\newcommand{\lsun}{\,L_{\rm \odot}}

\newcommand{\be}{\begin{eqnarray}}
\newcommand{\ee}{\end{eqnarray}}

\renewcommand{\arcmin}{^\prime}

\renewcommand{\farcs}{.\!\!^{\prime\prime}}

\def\etal{{et al.\thinspace}}
\def\eg{{e.g.,\ }} 

\newcommand{\benu}{\begin{enumerate}}
\newcommand{\eenu}{\end{enumerate}}
\newcommand{\bite}{\begin{itemize}}
\newcommand{\eite}{\end{itemize}}

\begin{document}

\shorttitle{Supernovae in Globular Clusters}
\shortauthors{Pfahl, Scannapieco, \& Bildsten}



\title{Globular Clusters as Testbeds for Type Ia Supernovae}

\author{Eric Pfahl\altaffilmark{1,2},  Evan Scannapieco\altaffilmark{3}, and Lars Bildsten\altaffilmark{1}}
\altaffiltext{1}{Kavli Institute for Theoretical Physics, University
of California, Santa Barbara, CA 93106.} 
\altaffiltext{2}{Current address: Institute for Defense Analyses, 4850 Mark Center Dr.,
Alexandria, VA 22311}
\altaffiltext{3}
{School of Earth and Space Exploration,  Arizona State University, PO
Box 871404, Tempe, AZ, 85287-1404.} 


\begin{abstract}

Fundamental mysteries remain regarding the physics of 
Type Ia supernovae (SNIa) and their stellar progenitors.  We argue here  that
important clues to these questions may emerge by the identification of those 
SNIa that occur in extragalactic globular clusters--stellar systems with well defined ages and 
metallicities.  We estimate an all-sky rate of $\approx 0.1 \eta
(D/100\mpc)^3\yr^{-1}$ for SNIa in globular clusters within a distance
$D$, where $\eta $ is the rate enhancement per unit mass as 
a result of dynamical production channels that are inaccessible in the
galactic field.  If $\eta\approx 2-10$, as suggested by observations and theory, the
combined efforts of accurate supernova astrometry and deep follow-up imaging should identify the
$\ga$1\% of nearby ($D<100$ Mpc) SNIa that occur in globular clusters. 

\end{abstract}




\section{Introduction}
\label{sec:intro}


Many questions remain about the most fundamental aspects of Type Ia
supernovae (SNIa), including the triggering and hydrodynamics of the
explosions \citep[e.g.,][]{Hillebrandt2000}, and the nature of their
stellar progenitors \citep[e.g.,][]{Yungelson2005}. The increasing
SNIa diversity, with some very bright \citep[e.g.,][]{Howell2006},
some very faint \citep[e.g.,][]{Kasliwal}, and some that do not
follow the Phillips (1993) relation
\citep[e.g.,][]{Jha} has energized the discussion of many 
possible formation scenarios.

While at most a few percent of white dwarfs explode as  SNIa
\citep[]{Pritchet}, there are only loose  constraints on the specific
binary evolution pathways \citep[e.g.,][]{Iben1984,Yungelson2005}.
There is a consensus that SNe Ia originate from thermonuclear ignition
and burning of a C/O white dwarf in a binary system.  Yet it remains
uncertain if the event is triggered by  accretion from a hydrogen-rich
companion or from a merger with another white dwarf \citep[see][]{Branch1995}, 
referred to as the single-degenerate and double-degenerate scenarios, respectively.
In the single-degenerate scenario, the main issues are the nature of
the companion, and if mass transfer can be sustained onto
the white dwarf at the favorable rate in a sufficient number of systems to match the
observed SNIa rate \citep{Maoz2008}. In the double-degenerate scenario
the primary issue is whether off-center carbon ignition can be avoided,
as this would transform the C/O white dwarf into a O/Ne/Mg white dwarf
\citep[e.g.,][]{Nomoto1985} rather than generate a SNIa
\citep[see however][]{Regos2003}.

In fact evidence is amassing that multiple progenitor scenarios lead to SNIa.
In particular, there is a clear disparity in the SNIa rates in late-type
galaxies with active star formation and elliptical
galaxies that contain mostly old (5--10\,Gyr) stars
\citep[]{Mannucci2005, Scannapieco2005,Sullivan2006}. This suggests
that $10^8-10^{10}\yr$ can elapse between the birth of the progenitors and the
explosions, a challenge for any single  progenitor scenario 
\citep[]{DellaValle1994,Mannucci2006}. Multiple routes to 
SNIa may well lead to diversity in the explosive outcomes. 
For example, low-luminosity SNIa are most prevalent in
early-type (i.e., E/S0) galaxies, while the most luminous events occur
only in star-forming galaxies \citep[e.g.,][]{Hamuy1996}.

 We propose that the study of SNIa in globular clusters (hereafter,
GCIa) may provide unique clues to understanding SNIa.  Although
globular clusters (GCs) and elliptical galaxies are both composed
mainly of old stars,  there are crucial differences between them.  
In a given GC, we are certain that all stars were
born within 1 Gyr of each other, whereas elliptical galaxies
often show evidence for a substantial spread of stellar ages
\citep[\eg][]{Trager2000}.  Secondly, in an individual GC, the stellar
metallicities are narrowly distributed, and thus the integrated
metallicity is a good measure of the metallicity of any of the
constituent stars. Until recently, there was only  one known exception
($\omega$ Cen) in the Milky Way \citep[]{Freeman1975, Bedin2004}, but
more recent work has found that  a few  massive ($>10^6 M_\odot$) GCs
have helium-rich sub populations \citep[e.g.,][]{Piotto2008}.
In any instance, GCs have metallicities low enough that a single GCIa
detection would place strong constraint on theoretical models
\citep[][]{Kobayashi1998,Hachisu1999,Piro2008}.

GCs are differentiated from the galactic field by  their high stellar
densities (often $\ga$$10^5\msun\pc^{-3}$) that trigger  frequent
close encounters between stars and binaries.  Such encounters are
responsible for the high incidence of exotic objects in GCs, including
X-ray binaries, rapidly spinning radio pulsars, blue stragglers  and
cataclysmic variables
\citep[\eg][]{Hut1992,Sills1999,Rasio2000,Pooley2006}.
Dynamics will almost certainly play an important role in the
production of GCIa progenitors \citep[][]{Shara2002,Ivanova2006,Rosswog2008},
likely increasing the  GCIa rate per unit mass.

We start in \S 2 by estimating the GCIa rate and discussing the
possible dynamical enhancements.  The observational challenges to
finding a GCIa are discussed in \S 3, where we motivate that the
maximum distance for such a search is 100 Mpc.  We also explain the
need for accurate astrometry of nearby SNe that will enable meaningful
followup observations.  We close in \S 4 by describing the
implications of detecting even a  single GCIa.


\section{The Supernova Rate in Globular Clusters}
\label{sec:gcsys}

Globular clusters (GCs) have $\sim$$10^5$--$10^6$ old ($>8$ Gyr) stars inside a 
few parsecs, with a wide range of $\lesssim 0.3 Z_\odot$ metallicities.
All galaxies contain GCs, with 
total numbers scaling as $\sim$100 GCs per $10^{10}\lsun$ \citep{Ashman}. 
The common measure of the GC number density is the $V$-band specific frequency,
\be 
S_N = N_{\rm GC} 10^{0.4({\cal M}_V + 15)}~, 
\ee
where $N_{\rm GC}$ is the number of GCs, and ${\cal M}_V$
is the absolute $V$-band magnitude of the galaxy \citep{Harris81}.  
Typical values of $S_N$ are
$\simeq$1 for spiral galaxies and $\simeq$2--5 for ellipticals \citep{Harris1991}. 
Though standard, $S_N$ is not the best choice for our purposes.  We are 
most interested in the fraction of stellar mass in GCs,
$F_{\rm GC} = M_{\rm GC}/M_g$, 
where $M_{\rm GC}$ is the total mass of the GC system and $M_g$ is 
the total stellar mass of the galaxy.  Given the galactic stellar mass-to-light ratio
$\Upsilon_V$, $F_{\rm GC}$ is related to $S_N$ by
\be
\label{eq:gcmassfrac} F_{\rm GC}  = 1.2\times 10^{-3} S_N m_5
\Upsilon_V^{-1} ~, 
\ee
where  $m_5 $ is the mean GC mass in units of $10^5 M_\odot,$ and  we use ${\cal M}_{V,\odot} = 4.8$ for the absolute magnitude of the Sun.  Photometric studies of GC systems find $m_5\simeq 2$, and old elliptical galaxies have $\Upsilon_V
= 3$, so that $F_{\rm GC} \approx 2\times 10^{-3}$ for most ellipticals, while it is somewhat less for spirals.
For some central dominant ellipticals at the centers of galaxy clusters, $S_N$ can reach 10 (Harris et al. 2009), corresponding to $F_{\rm GC} \approx  10^{-2}$.

\citet{Scannapieco2005} and \citet{Mannucci2005} proposed that the SN
Ia rate in a galaxy is the sum of two components, one proportional to
the total stellar mass $M_g$, the other proportional to
the star formation rate $\dot{M}_g$.  Such a model suggests 
that SNIa result from at least two evolutionary channels. 
For a particular galaxy, the
two-component rate can be written as
\be\label{eq:sbrate} {\rm SNR}(t) = A M_g(t) + B \dot{M}_g(t)~, \ee
where $A$ and $B$ are constants.  Using a well characterized sample of SNIa and host galaxies,
\citet{Sullivan2006} find basic agreement with eq.~(\ref{eq:sbrate})
and determine $A = (5.3 \pm 1.1)\times 10^{-14}\yr^{-1}\msun^{-1}$ and
$B = (3.9 \pm 0.7)\times 10^{-4}\yr^{-1}(M_\odot \yr^{-1})^{-1}$.

The value of $A$ is derived from SNIas  in E/S0 galaxies
with no discernible star formation (i.e., $B = 0$).  If these
galaxies are truly old, with negligible star formation in the past
$\approx$5--10\,Gyr, then a reasonable first guess is that the same
rate per unit mass also applies to GCs (the enhancement due to 
stellar dynamics is discussed below).  Adopting the galactic value of
$A$ for GCs, the GCIa rate in a galaxy is $A M_g F_{\rm GC}$.
An estimate of the local cosmic rate density of GCIas is obtained as
follows.

At low redshift, the total $K$-band luminosity density is  $j_{K}
\simeq (5\pm 0.5)\times 10^8\,L_{\odot K}\mpc^{-3}$ \citep[$H_0 =
70\kms\mpc^{-1}$;][]{Kochanek2001} and the fraction in E/S0 galaxies is
40--50\%.  Given the mean $K$-band stellar mass-to-light ratio
$\Upsilon_K \approx 1$, the contribution to the Ia rate density from
old stellar populations is $A j_K \Upsilon_K$.    Here we let
$\Upsilon_K = 1\pm 0.5$, averaged over all galaxy morphological types,
where the uncertainty largely reflects a range of model assumptions,
rather than measurement error
\citep[e.g.,][]{Cole2001}.  We find $A j_K \Upsilon_K
\approx (2.7 \pm 1.5)\times 10^{-5}\yr^{-1}\mpc^{-3}$.  This is
consistent with the recent measurement at 
$z\approx 0.1$ \citep[]{Dilday}, showing the dominance of the old stellar population for the local SNIa rate. 
The corresponding GCIa rate density is then $A j_K \Upsilon_K \langle
F_{\rm GC} \rangle$,  where $\langle F_{\rm GC} \rangle$ is the mean
GC mass fraction over a large number of galaxies.  If we adopt a
plausible value of  $\langle F_{\rm GC} \rangle = 10^{-3}$ (see
eq.~[\ref{eq:gcmassfrac}]), we estimate a 
\be\label{eq:ratemass}
\textrm{GC rate from mass alone} \approx 3\times 10^{-8}\yr^{-1}\mpc^{-3}.  
\ee
Under our given assumptions, we expect that the net uncertainty in this rate is a factor of $\approx$2.

 However, it  has been known for over 30 years that X-ray binaries 
 are $\ga$100 times more abundant per unit mass in GCs than
in the disk  \citep{Clark1975,Katz1975}.  Dynamical interactions involving single stars and
binaries occur frequently in GCs and naturally account for this
overabundance \citep[e.g.,][]{Bildsten2004}.   There are also
excellent observational and theoretical arguments that dynamics shapes
the GC populations of  blue stragglers, millisecond pulsars, and
cataclysmic variables \citep[e.g.,][]{Sills1999,Rasio2000,Pooley2006}.

Recently, \citet{Shara2002} and \citet{Ivanova2006} explored the idea
that SNIa progenitors can be formed by dynamical means in dense star
clusters, leading to a mean enhancement of the GCIa rate per unit
mass, $\eta$.  Shara \& Hurley (2002) suggest that the number of
WD-nondegenerate star binaries is similar in dense clusters relative
to the field, although they found strong differences in the masses of
the companion stars, which could be important in determining which
binaries support stable accretion.  The models described in Ivanova
\etal (2006) suggest an enhancement of $\eta \approx 1$-7 for
single-degenerate progenitors, where the range in $\eta$ reflects
variation with metallicity and other parameters.

In the double-degenerate case, Shara \& Hurley (2002) showed that
supra-Chandrasekhar WD-WD merger rate is over an order of magnitude
higher in dense clusters than in the field.  On the other hand,
\citet{Ivanova2006} suggest a more modest enhancement of $\eta \approx
2.$  Based on recent studies of the  prevalence of post classical
novae supersoft sources in M31 GCs, \citet{Henze2008} conclude that
the nova rate in GCs may be as much as ten times higher than in an old
field stellar population, and they suggest that GCIa may be detectable
in future surveys. Overall, it seems conceivable that $\eta \approx
1$-10 and that  observed GCIa may help differentiate  progenitors.

\section{Observational Considerations within 100 Mpc }
\label{sec:observations}

The maximum distance of interest is set by the need to find the underlying GC. GCs have a distribution of absolute magnitudes
given  by $dN/d M_V \propto \exp[-(M_V -
 M_{V,0})^2/2\sigma_V^2]$, where the dispersion is $\sigma_V \simeq
 1$--1.5 in relatively bright galaxies with $M_{V,{\rm gal}} <
 - 20$.  Over a wide range of host galaxy properties, the mean is
 $ M_{V,0} = -7.4$ to within a few percent
 \citep[e.g.,][]{Harris1991,Jordan2007} and has been detected at the 
100\,Mpc distance of the Coma cluster, where the
 turnover apparent magnitude ($\simeq$27.6) is accessible by the {\em
 Hubble Space Telescope} \citep{Harris2009}.
 Observations from the ground are presently limited to magnitudes of $m_V\la 26$,
 corresponding to the GC luminosity function turnover at distances of $\la$50\,Mpc.

For a given limiting apparent magnitude $V_{\max}$, the fraction of stellar
mass in GCs  brighter than $V_{\rm max}$ 
is
\be f(V< V_{\rm max}) = \frac{1}{\sqrt{\pi}} \int_{-\infty}^{(V_{\rm
max} - \widetilde{M}_V)/\sqrt{2\sigma_V^2}}  dx\ e^{-x^2}~, \ee
where $\widetilde{M}_V = M_{V,0} +{\rm DM} - 0.4(\ln 10) \sigma_V^2$, and DM is
the distance modulus.  When ${\rm DM} = 35$ and
$V_{\rm max} = 26$,  as $\sigma_V$ increases from 1 to 1.5, $f$
increases from 0.25 to 0.62, while the fraction of globulars with $V <
V_{\rm max}$ varies from 0.05 to 0.14.  Even though the fraction of
visible clusters may be small, the fraction of stellar mass contained
within these clusters can be substantial, and typically exceeds 50\%
when ${\rm DM} < 35$ and $\sigma_V$ takes its usual values of
$\simeq$1.3--1.5. Hence, observations of a galaxy at 100\,Mpc for which the GC census is
complete for clusters brighter than $V = 26$  finds $\simeq$15\% of
the globulars by number but $\simeq$ 60\% of the stellar mass.
This is an important point, since a dynamical  enhancement in the GCIa rate may 
favor more massive clusters, making it of considerable interest to pursue GCIa 
within ${\rm DM}=35$ \citep{Pooley2006}.  At distances greater than $\approx$100\,Mpc, it becomes extremely 
difficult to identify a significant number of GCs, and those detected in the outskirts of the galaxy 
will represent only a small fraction of the mass of the GC system.

From eq.~(\ref{eq:ratemass}) and our discussion of the dynamical enhancement of the 
GCIa rate per unit mass, we estimate a local GCIa rate of $\approx$$0.1 \eta \,(D/100\mpc)^3\yr^{-1}$. 
 A plausible value of $\eta \sim 10$ results in $\approx$1
GCIa per year within 100 Mpc, which is about $1\%$ of the total SNIa rate within 100
Mpc. What are the prospects for carrying out such a search?  The
typical SNIa within 100 Mpc would have a peak visual magnitude of
$m_V\approx 16-17$ and even the subluminous, 1991bg-like SNIa would be
found at these distances in the upcoming wide angle (one-tenth of the
sky) nearby SNe surveys (e.g., Palomar Transient Factory, SkyMapper and Pan-Starrs1). The
SNIa yields from these new surveys, as well as the increasing numbers
from targeted galaxy and cluster searches by LOSS-KAIT, CHASE, and
ROTSE, makes the time right for explicit GC identification efforts. In
some cases, prior HST or ground-based studies will have GC catalogs to
cross-list locations. However, the typical case will require 
waiting until the SNIa has faded below the GC light.

The few well-studied late-time SNIa light curves reach $M_V\approx -7$
after $\approx 600$ days (e.g., Sollerman et al. 2004; Lair et al. 2006),
at which point their fade rates are
$1.4$ magnitudes per 100 days in $BVR$ (Sollerman et al.  2004) with
colors of $V-R\approx -1$ and $B-V\approx 0$, much bluer than a
GC. The $I$ band decays more slowly  ($\approx 1$ mag in 100 days),
again pointing to $BVR$ for discovery.  One possible way to first
discern the presence of an underlying GC would be the 
detection of a modified $BVR$ color evolution as the redder 
$BVR$ colors of the GC  ($V-R\approx 0.4$-1) begin to shine through.

After identifying a candidate GCIa, high angular resolution observations  are required: 
(1) to confirm that the SNIa and GC lie along the same sightline, and (2) 
to minimize the likelihood that the SNIa occurred in the host galactic field, in front or behind the GC.
If the GC and SNIa are found to overlap within a resolution element of diameter $\theta$ (in arcsec)  
the probability that the SNIa occurred in the field is roughly 
$L_\theta/L_{\rm GC} \equiv \epsilon,$  where $L_{GC}$ is the GC luminosity and $L_\theta$ is the luminosity in 
field stars within the resolution element.  A definitive GCIa detection requires a small value of $\epsilon,$ and a correspondingly 
small value of $L_\theta$.  Since surface brightness, and thus $L_\theta$, generally falls with increasing galactocentric radius, the strongest GCIa 
candidates will be located far from the centers of their host galaxies.  To illustrate this point more quantitatively, 
we assume a de Vaucouleurs profile with a surface brightness
$\mu_e = 19.5\,\textrm{mags/arcsec$^2$}$ at the half-light radius,
$R_e$ \citep[e.g.,][]{Djorgovski1987}.  We further assume that the target GC has magnitude 
$M = -7.5$ and that the GC is unresolved (the typical GC half-light diameter 
is $\la$$0\farcs1$ beyond 10\,Mpc).  With these assumptions, we find that the radius at which $L_\theta/L_{\rm GC} = \epsilon$ is 
given by 
\be
\frac{R}{R_e} = \left[
\frac{34}{25} + \frac{3}{10}\log\left(\frac{D_{10}^2\theta^2}{\epsilon}\right)
\right]^4~,
\ee
where $D_{10} \equiv D/10\mpc$.  For $\epsilon = 0.1$ and  $D_{10} =
\sqrt{10}$ ($\simeq$31.6\,Mpc) we find $R/R_e \simeq$$3.4$ when
$\theta = 0.1$ arcsec, and $R/R_e \simeq$$0.78$ when $\theta =
0.02$ arcsec.  At a distance of 100\,Mpc, the same two $\theta$
values give $R/R_e \simeq 7.6$ and $\simeq$ 2.3.   However, at 1.0
arcsec resolution, $R/R_e \simeq 14.7$ even when $D_{10} =
\sqrt{10}$. Since typical half-light radii are $R_e = 1$--4\,kpc,
it is clear that $\la$$0\farcs1$ resolution is required to achieve
modest $R$ for small $\epsilon$, which can only be accomplished from
space or with ground-based adaptive optics.  Even then, the best GCIa
candidates will be at $R > 10\kpc,$ which  requires accurate astrometry
of both the active SNe and the possible underlying
GC.  These limits highlight the value of extremely accurate astrometry
for making GCIa measurements in the future with large ground-based telescopes.


\section{Implications of a Discovery}
\label{sec:implications}

A major open question is how a 10\,Gyr old stellar population produces an appreciable
SNIa rate, as  this requires double-degenerate mergers or stable
accretion from relatively low-mass donors, neither of which are 
currently-favored for SNIa production.  While some elliptical galaxies
show definite signatures of relatively recent  low-level star
formation within an otherwise very old system
\citep[\eg][]{Trager2000}, we can be confident that no
new stars are forming in old GCs.  A single, definitive GCIa detection would 
demonstrate that, in fact, SNIa do occur in truly old stellar systems. 

Secondly, any systematic trends in GCIa properties with metallicity
contain information about the physics of the explosions.  GCs are,
with few exceptions, extremely uniform in their chemical
compositions, and the same cannot be said of elliptical galaxies
\citep[\eg][]{Mehlert2003}.  Of course, the metallicity
can vary a great deal between GCs, but the metallicity of an
individual cluster is readily determined from its photometric colors.  The detection of a
single [Fe/H] $< -1$ GCIa would place strong constraints on models
of stable white-dwarf accretion from a non-degenerate companion
\citep[\eg][]{Kobayashi1998,Hachisu1999}, and analysis of a handful of GCIa would strongly 
constrain models of SNIa lightcurves and pre-explosion simmering \citep[\eg][]{Timmes2003,Piro2008}.

A potential complication is that the dense stellar environment of the GC may open
up exotic paths to SNIa (e.g., Rosswog et al. 2008). We would hope that such an outcome 
would be revealed in comparison between the GCIas and the field SNe, 
both  in rates \citep[\eg][]{Shara2002,Ivanova2006} and systematic properties,
such as lightcurve shapes.  If the rates are sufficiently high, GCIa 
may prove decisive in implicating nonstandard Ia progenitors such as
double-degenerate mergers.

Because it takes approximately two years for a typical Ia to fade to
the luminosity  of an average globular cluster, GCIa will require late-time observations. 
This is not common practice, although it would
require minimal investment of telescope time for the
closest supernovae, as the total rate is only $\approx 10\yr^{-1}$ within
30 Mpc. Accurate astrometry of these events is also critical to the later GC search. 
For the moment, we must appeal to the Ia archives.  In the Sternberg
catalog\footnote{\url{http://www.sai.msu.su/sn/sncat/}}, there are 112 SNIa 
identified from 2005 to 2007 within $z \leq 0.025$ (7500 km/s). Of these, 34 
are hosted by E/S0 galaxies, and 6 in this subset are
separated by more than $1\arcmin$ from the centers of their hosts.
Only careful follow-up observations will tell if the first GCIa has
already been detected.


\acknowledgments

This work was supported by the National Science Foundation under
grants PHY 05-51164, AST 07-07633, and AST 08-06720.   We thank James 
Rhoads and the anonymous referee for helpful comments.
  
  
\bibliographystyle{apj}


\fontsize{10}{10pt}\selectfont

\bibliographystyle{apj}

\end{document}